# Title: Transitions between homophilic and heterophilic modes of cooperation

Authors: Genki Ichinose[1]*, Masaya Saito[1], Hiroki Sayama[2], and Hugues Bersini[3]

**Affiliations:**

[1] National Institute of Technology, Anan College, 265 Aoki Minobayashi, Anan, Tokushima 774-0017, Japan.

[2] Collective Dynamics of Complex Systems Research Group, Binghamton University, State University of New York, Binghamton, NY 13902-6000, USA.

[3] Université Libre de Bruxelles, 50, av. Franklin Roosevelt 1050, Brussels, Belgium

*Correspondence to: ichinose@anan-nct.ac.jp

**Abstract**: Cooperation is ubiquitous in biological and social systems. Previous studies revealed that a preference toward similar appearance promotes cooperation, a phenomenon called tag-mediated cooperation or communitarian cooperation. This effect is enhanced when a spatial structure is incorporated, because space allows agents sharing an identical tag to regroup to form locally cooperative clusters. In spatially distributed settings, one can also consider migration of organisms, which has a potential to further promote evolution of cooperation by facilitating spatial clustering. However, it has not yet been considered in spatial tag-mediated cooperation models. Here we show, using computer simulations of a spatial model of evolutionary games with organismal migration, that tag-based segregation and homophilic cooperation arise for a wide range of parameters. In the meantime, our results also show another evolutionarily stable outcome, where a high level of heterophilic cooperation is maintained in spatially well-mixed patterns. We found that these two different forms of tag-mediated cooperation appear alternately as the parameter for temptation to defect is increased.

**Kerwords:** evolution of cooperation, tag, spatial structure, migration, segregation

**Main Text:**

**Section: Introduction**

Evolution of cooperation has attracted attention of both biological and social scientists. Cooperation benefits others while incurring cost to the actor. In contrast, defection allows the actor to receive the benefit without paying any cost. Thus, natural selection and social mimetism tend to naturally favor defection in a well-mixed population. It is known that some kind of mechanism to facilitate the regrouping of cooperators, either in a temporal sequence (tit-for-tat) or in a spatial distribution, is needed for cooperation to evolve (Nowak 2006). Among them, kin selection (Hamilton 1964) has the longest history. According to this theory, altruistic genes can survive and spread into a population by helping relatives who share the same genes despite the sacrifice of individuals. A phenotypical tag can be regarded as an indicator of such relatedness although it is a weak and initially arbitrary characteristic. The effect of such a tag has been studied in the context of the evolution of cooperation. Riolo et al. (2001) have shown how the existence of tags strongly promotes cooperation in one shot PD even in the absence of spatial structure, where every agent helps another agent if the distance between their tags is less than a given threshold. Their model assumed that agents with similar tags always cooperate, and therefore cooperation was greatly enhanced in their model. Roberts and Sherratt (2002) showed that, without this assumption, cooperation no longer evolves in the absence of spatial structure. In contrast, analytical studies showed that cooperation can be enhanced by a tag mechanism even without spatial structure, but it requires complicated conditions (Antal et al. 2009; Traulsen & Nowak 2007; Traulsen & Schuster 2003).

Some studies showed that the presence of a spatial structure can help cooperation to spread when combined with the tag mechanism (Axelrod et al. 2004; Hammond & Axelrod 2006; Hartshorn et al. 2013; Jansen & van Baalen 2006; Shutters & Hales 2013; Spector & Klein 2006; Traulsen & Claussen 2004). Intuitively, restricting cooperation to agents sharing the same tag decreases the exposure to defectors (compared to unconditional cooperation) and increases the frequency of cooperation (compared to a cooperation restricted to the opposite tag). For instance, Spector and Klein (2006) showed in a one-dimensional population structure how cooperation evolves in the presence of the tag mechanism. Other models revealed the same fact by assuming a two-dimensional population structure (Axelrod et al. 2004; Hammond & Axelrod 2006; Hartshorn et al. 2013; Jansen & van Baalen 2006; Shutters & Hales 2013; Traulsen & Claussen 2004). Especially, Traulsen and Claussen (2004) observed a strong segregation between agents with different tags.

In all of those previous studies reviewed above, agents were embedded in a spatial structure *a priori* and forced to play games with their local neighbors without any active mobility. In contrast, here we propose a new model in which we let agents to migrate spatially under certain conditions, and allow the strategy for migration to evolve together with the tag-based strategies for game play. This model modification is inspired by the recent finding that contingent migration generally promotes cooperation (Aktipis 2004; Chen et al. 2012; Helbing & Yu 2009; Ichinose et al. 2013; Jiang et al. 2010; Roca & Helbing 2011). To our knowledge, the effect of such migration on tag-mediated cooperation in spatial models has never been discussed before, except in our recent study (Bersini 2014). Furthermore, in most of the earlier tag models, the semantic role of the tags was always fixed beforehand (e.g., an agent cooperates with another agent whose tag is identical or similar), although it would be more natural to assume that such a role should also arise as the result of evolution. The model we propose assumes that the semantic role of tags is subject to evolution, together with the tag threshold for migration and strategies for game play. With this model, we aim to find necessary conditions for the emergence of cooperation and segregation, and also to show how cooperation and segregation co-evolve hand in hand.

**Section: Model**

We developed an agent-based model in which individual agents are distributed over a two-dimensional square lattice and play the PD game with their neighbors. The square lattice is composed of $L \times L$ sites with periodic boundary conditions. Each site is either empty or occupied by one agent. Agents can migrate to empty sites which represent spatial regions. Initially, agents are randomly distributed over the square lattice. There are two tags (green and red) for the agents. Half of the randomly selected agents are assigned green and the other half are assigned red. As for the PD game, each agent has two cooperation levels, one is cooperation probability, $p_{Cs} \in [0, 1]$, to the opponent whose tag is the same as the focal agent, and the other is that probability, $p_{Cd} \in [0, 1]$, to the opponent whose tag is different from the focal agent. As the initial setting, an equal number of agents which respectively have $(p_{Cs}, p_{Cd}) = (1, 0)$ or $(p_{Cs}, p_{Cd}) = (0, 1)$ is distributed over the space unless otherwise noted. Thus, at the beginning of the simulation half of the agents always cooperate with the identical tag agents and the other half always cooperate with the different tag agents. Moreover, each agent is allowed to migrate to another site depending on the tag states of the neighbors. The preference for the tag is represented by $\eta \in [-1, 1]$, which is randomly assigned as the initial setting. $\eta = -1$ means the agent is completely heterophilic i.e. the agent is satisfied when surrounded by agents with the other tag. $\eta = 1$ means the agent is completely homophilic i.e. the agent is satisfied when surrounded by agents with the same tag. Thus, $\eta = 0$ means that the agent has no preference about the tag. Therefore, each agent has three genetic values $p_{Cs}$, $p_{Cd}$, and $\eta$, which are all subject to evolution.

The population density is given by $\rho$ (i.e., the fraction of empty sites is $1 - \rho$). Thus, the population size is represented by $N = L \cdot L \cdot \rho$. The population density remains constant throughout a simulation run, since agents will never die or born.

Agents are updated asynchronously in a randomized order. The algorithm for updating an agent consists of the following three phases:

1. Game play.

A randomly selected agent plays the PD game with its neighbors (within the Moore neighborhood) and accumulates the payoffs resulting from the games. If there are no other agent within the neighborhood, no game is played. In each game, two agents probabilistically decide whether to cooperate or defect simultaneously based on their current strategies. They both obtain payoff $R$ for mutual cooperation while $P$ for mutual defection. If one selects cooperation while the other selects defection, the former receives the sucker's payoff $S$ while the latter receives the highest payoff $T$, the temptation to defect. The relationship of the four payoffs is usually $T > R > P > S$ in PD games. Following the parameters setting used in the model by Nowak and May (1992), we used $P = 0$, $R = 1$, and $S = 0$, while $T = 1 + b$ ($b > 0$), and $b$ was varied as a key experimental parameter. Since the cooperation levels of each agent are defined by continuous values, the following expected payoff, $p_i$, is used instead of the actual payoff of the agent $i$, playing with an agent $j$.

$p_i = (1 + b - b \cdot p_{Csi}) \, p_{Csj}$    if $(Tag_i = Tag_j)$ and           (1)

$p_i = (1 + b - b \cdot p_{Cdi}) \, p_{Cdj}$    $(Tag_i \neq Tag_j)$

2. Strategy updating.

After the randomly selected agent plays the PD game with its neighbors, the neighbors also play the game with their own neighbors. Once all the games, including the neighbors' games, have taken place, the focal agent imitates the strategy of the agent that achieved the highest total payoff among its neighbors, including itself (if there is a tie one agent is randomly selected). At the strategy updating, a small mutation occurs to the original three values. The new values are picked up from the Gaussian distribution where the means are the original $p_{Cs}$, $p_{Cd}$, and $\eta$ values and the standard deviation is $\sigma_s$ for the strategies and $\sigma_m$ for $\eta$. If there are no other agents within the neighborhood, the agent inherits its original strategies with the mutations.

3. Migration.

The agent decides to move or not depending on parameter $\tau$. $\tau$ represents the threshold for the migration and is defined as follows: $\tau = 1 - |\eta|$. As described above, the positive and negative $\eta$ means the different preference for the tag. Thus, there are two different criteria for the migration. Each agent moves to a randomly selected empty site if the following condition is satisfied.

$\tau_i < n_d / (n_s + n_d)$    if $(\eta > 0)$ and      (2)

$\tau_i < n_s / (n_s + n_d)$    $(\eta < 0)$

where $n_s$ ($n_d$) is the number of the neighbors that have the same (different) color as the agent $i$.

If there are no other agents within the neighborhood, the agent stays at the same site.

We regard $N$ time steps as one generation, in which all agents are selected once, on average, for the above three phases. The parameters set used in the simulations is $L = 50$, $\rho = 0.9$, and $\sigma_s = 0.0001$, $\sigma_m = 0.05$ and $b$ is varied unless otherwise noted. The model code is available at https://www.openabm.org/model/4635/version/1/view.

**Section: Results**

**Subsection: Evolution of cooperation against the temptation to defect**

We conducted computer simulations of the agent-based model described in the Model. Each simulation was run for 10,000 generations, and the results were collected from the last 1,000 generations unless otherwise noted. We conducted 100 independent simulation runs for each experimental condition, and used their averages as the final results. The segregation level was quantified by calculating

$s = (1/N) \sum_{i \in N} (n_s / n_i)$,     (3)

where $N$ is the number of agents, $n_i$ the number of nearby agents around individual $i$, and $n_s$ the number of such neighbors whose tag is identical to $i$'s tag. If $n_i$ is 0, then $n_s / n_i$ is defined as 0. As described in the Model, each agent has two real values as interaction strategies. $p_{Cs}$ is the cooperation probability to the opponent whose tag is the same as the focal agent while $p_{Cd}$ is the cooperation probability to an agent with the different tag. In the results, the two values for all agents are averaged. Since we use continuous strategy sets for $p_{Cs}$ and $p_{Cd}$ rather than discrete strategies, the agents cannot be described simply as "cooperator" or "defector" any more. Instead, they can be described to have a "high cooperation level" when their values of $p_{Cs}$ and $p_{Cd}$ are close to 1. In contrast, they are described to have a "low cooperation level" when these values are close to 0. Hereafter, we use these terms in the results.

The key finding we obtained from the simulations is that agents with high cooperation levels can self-organize in two different ways, depending on the parameter for temptation to defect, which we call $b$. Figure 1 shows the evolution of cooperation levels ($p_{Cs}$ and $p_{Cd}$) as a function of $b$ together with the screen shots of simulation runs obtained at their final generations. At the beginning of the simulations, the two game play strategies, $(p_{Cs}, p_{Cd}) = (1, 0)$ and $(p_{Cs}, p_{Cd}) = (0, 1)$, were equally present in the population. When $b = 0$, there is no incentive for agents to defect, while there is still an incentive for cooperation to avoid defection (low cooperation levels) because playing the game with agents with low cooperation levels would yield no payoff. Therefore, there is an evolutionary pressure for agents with high cooperation levels to form clusters. For such agents with high cooperation levels, it is structurally easy to form clusters with agents with the same tag agents rather than doing so with agents with the different tag at $b = 0$. This is because, as shown next, clustering with the different tag agents needs more complex spatial arrangements of agents. Therefore the strategy $(p_{Cs}, p_{Cd}) = (1, 0)$ simply wins when $b = 0$ (Fig. 2(A)). As $b$ is increased ($0 < b \leq 0.26$), agents with low cooperation levels increase their dominance by exploiting the same tag clusters that were seen at $b = 0$. As a response, agents with high cooperation levels try to avoid the exploitation by cooperating with agents with a *different* tag. When their evolutionary strategy reaches the point to fully cooperate with agents with a different tag ($b=0.26$), a unique, "maze-like" spatially well-mixed pattern arises (Fig. 2(B)) as seen in the earlier study (Traulsen & Claussen 2004). As $b$ is further increased ($0.26 < b \leq 0.42$), a more complex evolutionary dynamics emerges. In this parameter regime, agents with high cooperation levels initially try to cooperate with agents with a different tag, but since this strategy cannot completely eliminate defective behaviors, the agents with high cooperation levels "switch" their strategies in the middle of evolutionary processes (see the evolutionary trajectory in Fig. 2(C)) to cooperate with agents with the same tag as theirs, resulting again in the formation of segregated patterns. Once such segregated patterns are established, agents with low cooperation levels can survive only at the edges of clusters of cooperators (see the behaviors in Fig. 2(C) ($b = 0.34$)). When the temptation to defect is too large ($b > 0.42$), there is no merit for cooperation to cluster any more, resulting in random spatial patterns (Fig. 2(D)).

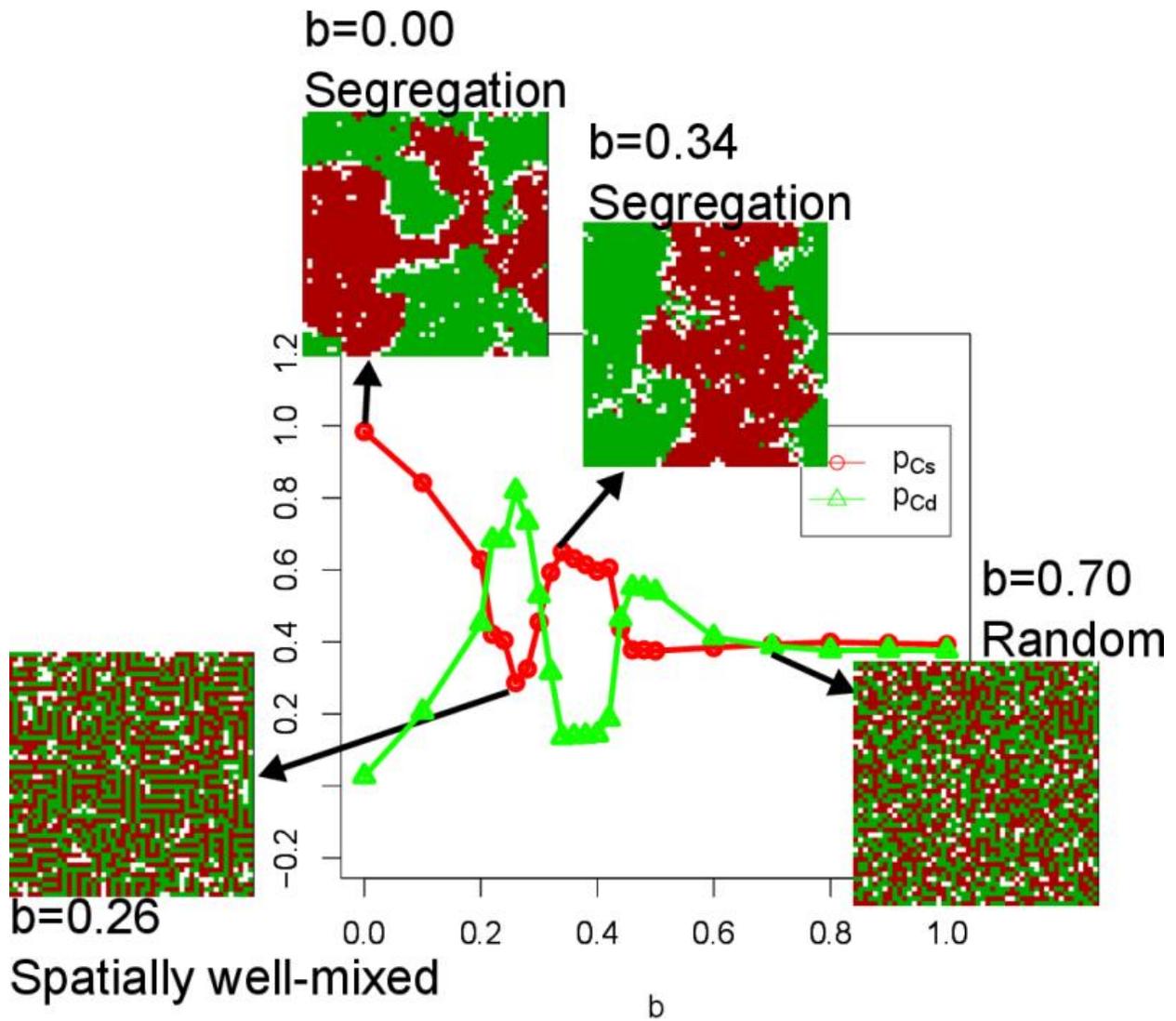

**Fig. 1.** Evolution of cooperation levels ($p_{Cs}$ and $p_{Cd}$) plotted against *b* together with the screen shots obtained at the 3,000th generation. See also Fig. 2 for visualization of cooperation levels and their evolutionary trajectories.

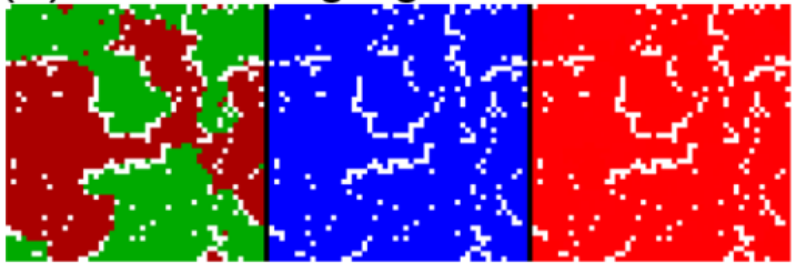
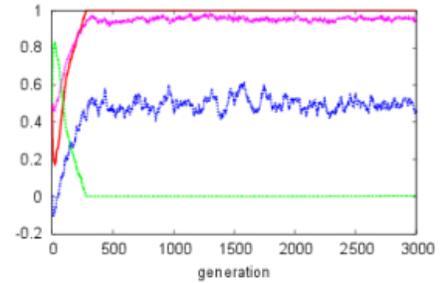
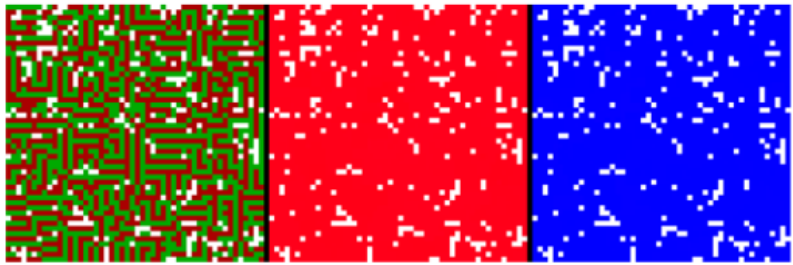
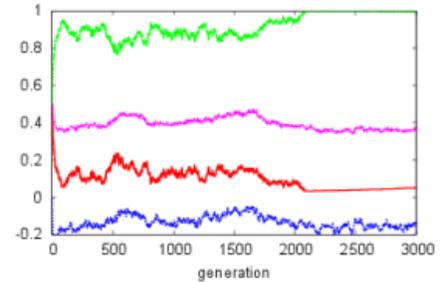
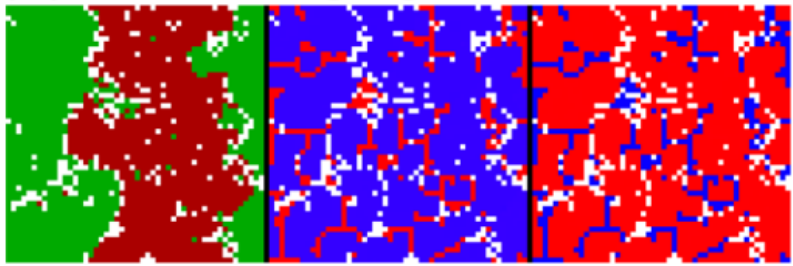
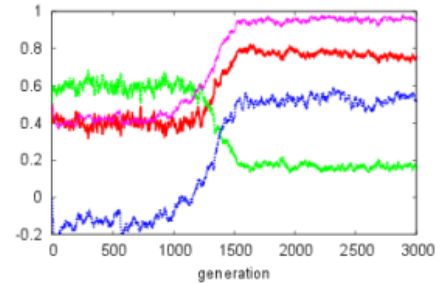
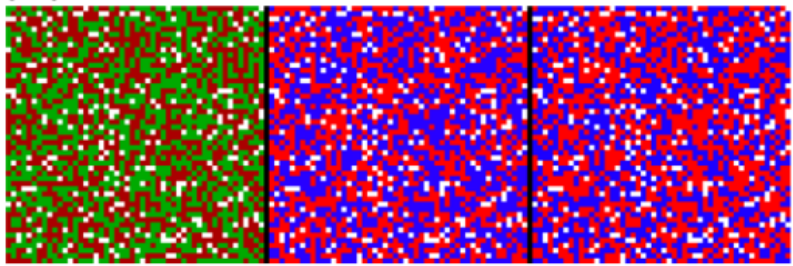
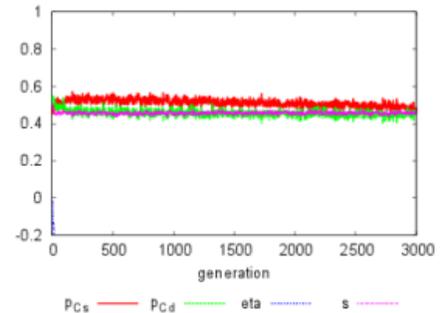

**Fig.2.** The screen shots at the final generation (3,000) and their evolutionary trajectories of each parameter for different $b$ values. In the screen shots, the left panel corresponds to the tags (green and red), the center panel corresponds to $p_{Cs}$, and the right panel corresponds to $p_{Cd}$. The blue corresponds to 1, the red corresponds to 0, and the intermediate values are represented by the mixed color.

In these results presented above, the temptation to defect $b$ was the only parameter being varied. As seen in Fig. 1, cooperation survives even when the temptation to defect is quite high ($p_{Cs} > 0.39$, $p_{Cd} > 0.37$ at $b = 1.0$). This is due to the spatial effect that allows cooperative agents to cooperate with each other locally to sustain themselves (regardless of their tags). Figure 3 shows the segregation level $s$ and the tolerance to the tag composition of the neighborhood $\eta$ when varying $b$. The two outcomes, i.e., segregation and spatially well-mixed, can be clearly distinguished.

We also notice that these two outcomes can be obtained even for the same $b$ value, depending on the initial distributions of agents. For example, with $b = 0.1$, segregation was obtained 91 times out of 100 simulation runs.

The other property $\eta$ (the tolerance to the tag composition of the neighborhood) also evolves differently depending on $b$ values. In all segregated outcomes, $\eta$ was above 0.4 at least, meaning that each agent moves to another empty site when surrounded by more than 60% of agents with a different tag, on average. This result has an interesting similarity with the seminal work of Schelling (1971), in which he observed that even a small value of $\eta$ (around 0.33) could eventually lead to a segregated state (even though his original model was not an evolutionary one). Our results indicate that, when segregation occurs, $\eta$ tends to converge towards a similar value as the critical value obtained by the Schelling's segregation model.

When spatially well-mixed states arise, $\eta$ evolves around −0.15, meaning that each agent moves when surrounded by more than 85% of agents with the same tag as its own, on average. This indicates that, in such cases, agents have little incentive in moving and definitely prefer to stay in a very heterogeneous neighborhood.

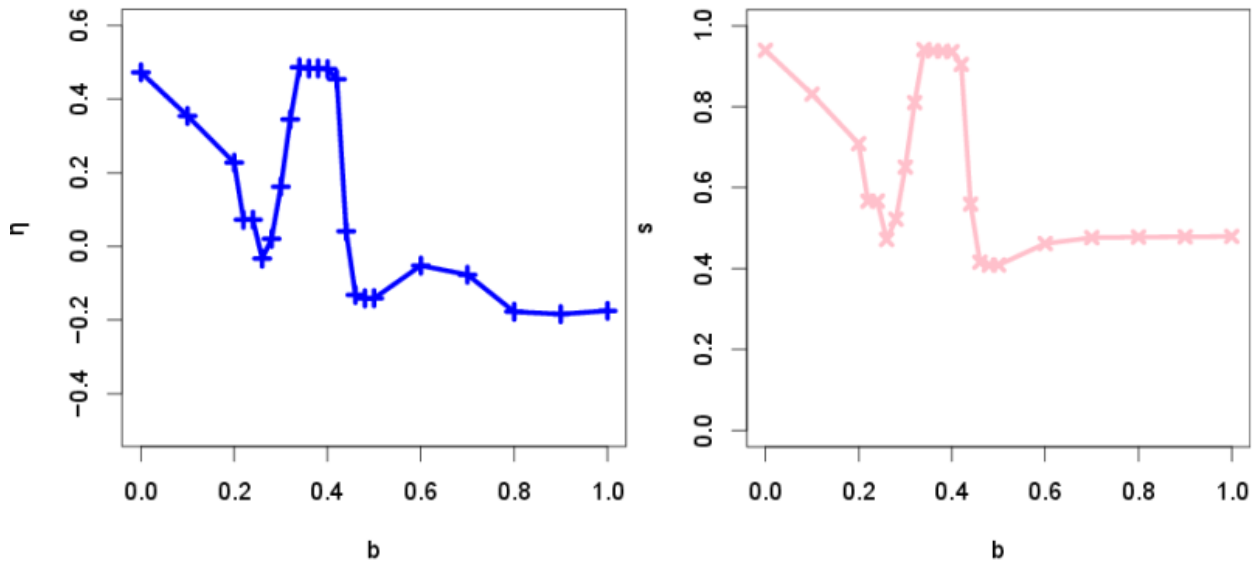

**Fig. 3.** Evolution of the segregation level $s$ and the tolerance to the tag composition of the neighborhood $\eta$ against $b$.

As in the Nowak's original spatial evolutionary game (Nowak & May 1992), the defection pressure parameter *b* appears to be quite critical for clustering of cooperators. For small values, all agents cooperate, in contrast with the high values for which they all defect. The most interesting and unexpected phenomena such as clusters of cooperators can be observed for the intermediary values, as also shown in our model (in addition we also study the effect of tag and migration).

**Subsection: Comparison with unconditional migration**

One important question to be asked is how effective the tag-based migration is in promoting the evolution of cooperation compared to an unconditional, or tag-blind, migration. To address this question, we conducted another set of simulations. Figure 4(A) shows the results obtained using a revised model with agents migrating in a random fashion (irrespective of the tags) named unconditional migration ("UM") and compares it with our previous simulation results (labeled "TM", for tag-based migration, shown as horizontal reference lines in the plots), with *b* fixed to 0.1. For the unconditional migration cases, the probability of unconditional migration was varied from 0 to 1. To compare UM and TM fairly, the expected cooperation level, $p_C = p_{Cs} \times s + p_{Cd} \times (1 - s)$, is used since the encounter to other agents with same or different tags is different in each model (See Fig. 4(B)). Cooperation was slightly enhanced in the unconditional migration case for a small migration probability, but it is well below the result of the tag-based migration case. Overall, tag-based migration is shown to be much more effective than unconditional migration in promoting the evolution of cooperation.

Focusing on the small probability of unconditional migration (0.0 to 0.3) (Fig. 4(B)), $p_{Cd}$ is much higher than $p_{Cs}$. This indicates that the strategy $(p_{Cs}, p_{Cd}) = (0, 1)$ always wins in this range. This is because, with *b* = 0.1 in the unconditional migration, encountering an agent with a different tag occurs more frequently than encountering an agent with a same tag. Therefore, cooperation is enhanced by interacting with agents with a different tag.

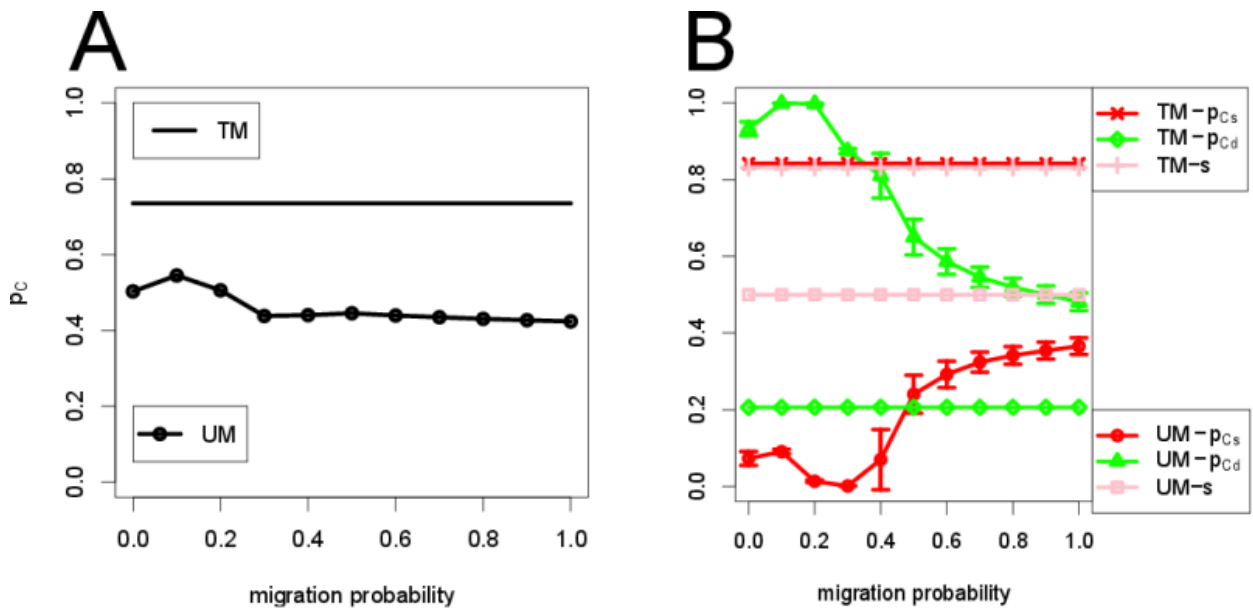

**Fig. 4.** (A) Comparison with unconditional migration (labeled "UM"). "TM" is tag-based migration. In all cases, $b = 0.1$. $p_C$ of "TM" is much higher than for the random migration. (B) $p_{Cs}$, $p_{Cd}$, and $s$ in UM and TM are shown, respectively. Tag-based migration can maintain high level of cooperation by cooperating with the identical tag agents more frequently. Compare the red line marked with crosses (TM) and the red line marked with circles (UM). The vertical bars of UM indicate standard deviations.

**Subsection: Robustness of cooperation and segregation**

To study the effect of the other model parameters, we also varied the other three parameters one by one while keeping the rest constant. Figure 5(A) shows the dependence of cooperation and segregation levels on the population density $\rho$ ($0.10 \leq \rho \leq 0.99$). When the space is relatively sparse (e.g., $0.3 < \rho < 0.7$), it is difficult for agents to form clusters. Moreover, when $\rho$ is too low (e.g., $\rho = 0.1$), agents simply cannot find each other, and hence interaction among agents and migration hardly occur. Such sparseness prevents the appearance of identical tag groups. Therefore, the high population density (e.g., $0.80 \leq \rho \leq 0.99$) is necessary for high cooperation and high segregation to take place.

Figure 5(B) shows the dependence of cooperation and segregation levels on the mutation rate of strategies $\sigma_s$ ($10^{-5} \leq \sigma_s \leq 10^{-1}$). As $\sigma_s$ increases, segregation levels slightly decrease whereas the cooperation with the different tag ($p_{Cd}$) increases. This is because each strategy needs to be stably linked to each tag in order to make the tag-mediated cooperation to work out effectively. Therefore, the strategies must be mutationally stable enough to maintain the homogeneous tag groups in which agents cooperate with each other.

Figure 5(C) shows the dependence of cooperation and segregation levels on the mutation for tag threshold $\sigma_m$ ($0.01 \leq \sigma_m \leq 0.20$). When $\sigma_m$ is large ($0.09 \leq \sigma_m \leq 0.20$), migration occurs too frequently, resulting in the instability and destruction of homogeneous groups. Therefore, the tag threshold must also be mutationally stable to maintain homogeneous tag groups.

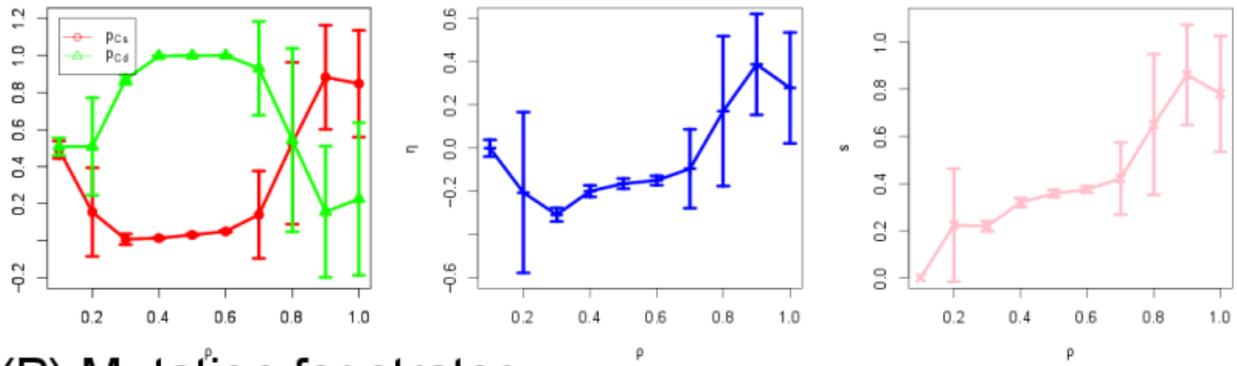
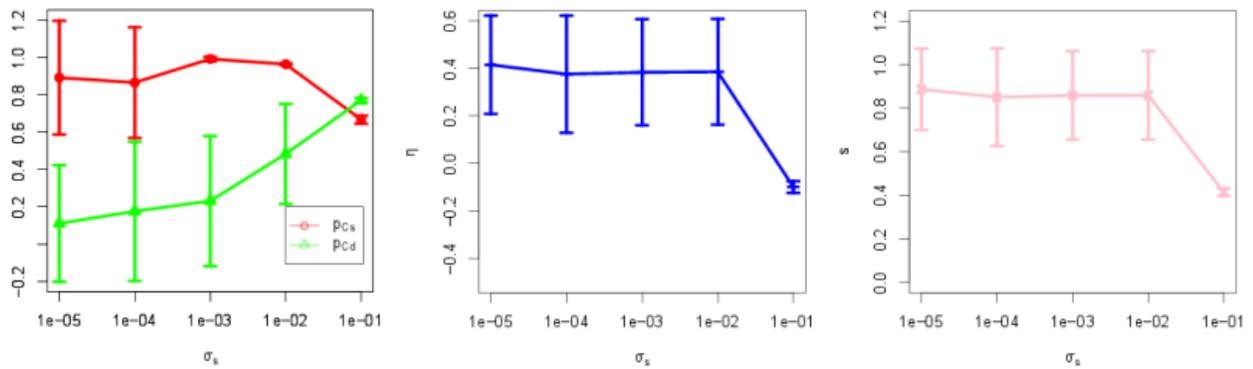
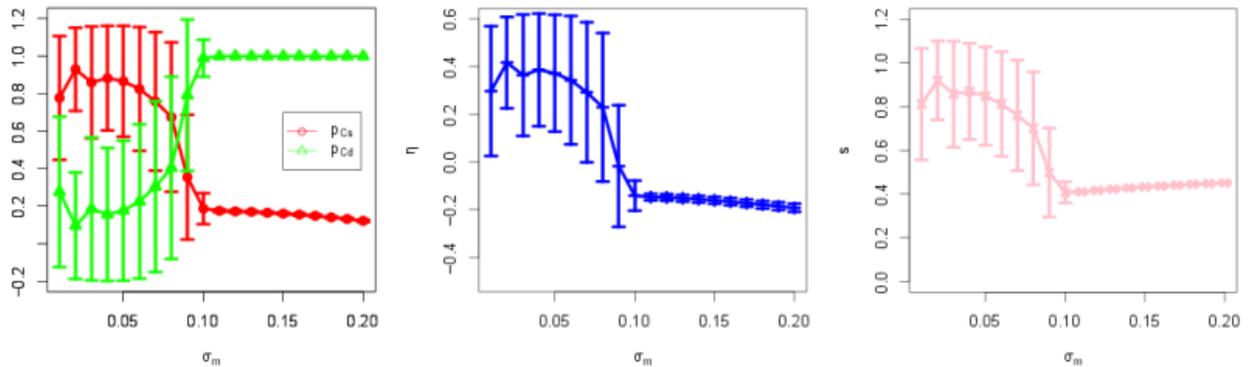

**Fig. 5.** Sensitivity analysis. Each of the three parameters ($\rho$, $\sigma_s$, and $\sigma_m$) is varied while keeping the other parameters constant. The vertical bars indicate standard deviations. $b = 0.1$.

Finally, in order to explore the robustness of segregation, we have extended the original model in two ways: increasing the number of tags and enlarging the neighborhood.

**Subsection: Increasing the number of tags**

While the number of tags was 2 in our original model, this number was increased to 4, allowing more "tribes" to emerge in the model. Figure 6 shows the $p_{Cs}$, $p_{Cd}$, and $s$ against $b$, for three different numbers of tags (2, 3 and 4). A final screen shot of a simulation with $b = 0.3$ and 4 tags is also shown. Interestingly, the intermediate $b$ values ($0.3 \leq b \leq 0.5$) yield higher segregation $s$ for three- and four-tag cases than the original two-tag case. This is because the possibility that agents meet the different tag agents increases as the number of tags increases. When cooperation is dominant ($0.0 \leq b \leq 0.2$) and in the presence of more tags, cooperating with the different tags is more beneficial to the agents. However, once defection becomes more attractive ($0.3 \leq b$), the only way for cooperation to survive is by favoring clusters composed of identical agents. Thus, communitarian cooperation and segregation co-evolve hand in hand in the range $0.3 \leq b \leq 0.5$. For further larger values of $b$ ($0.6 \leq b$), defection becomes extremely attractive destroying easily any attempt of cooperative cluster. In the two-tag cases, clustering with the same tag easily takes place compared to the different tag when $b$ is small. In contrast, as the number of tags is increased, since the possibility to meet the different tag is also increased, ($p_{Cs}$, $p_{Cd}$) = (0, 1) simply wins in the three-tag and four-tag cases. Therefore, segregation for the small $b$ is not observed and is only observed for the intermediate value of $b$ ($0.3 \leq b \leq 0.5$) to prevent clusters from defection invasion in those cases.

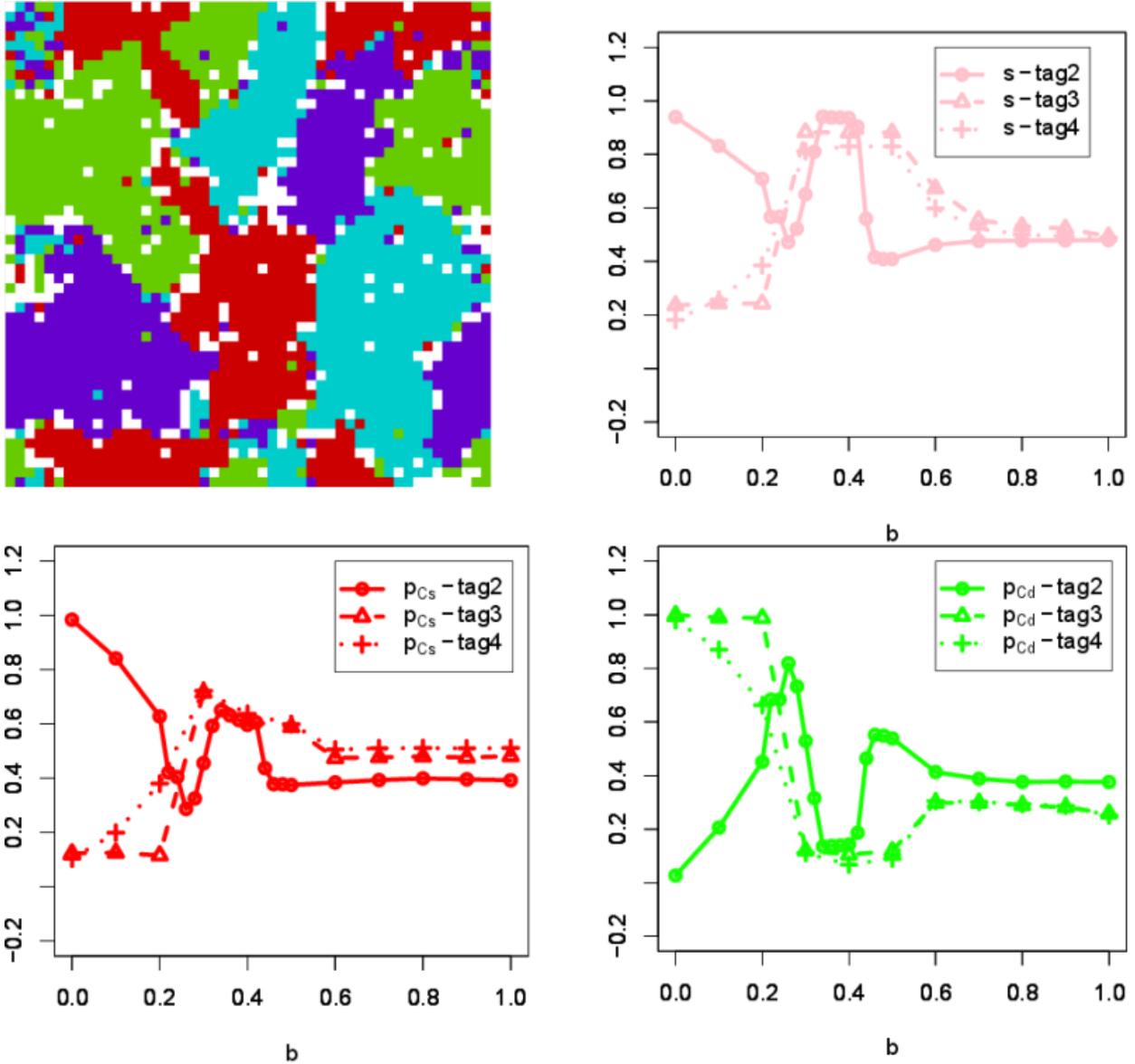

**Fig. 6.** Increasing the number of tags. $p_{Cs}$, $p_{Cd}$, and $s$ for the different tag number are shown. The screen shot shows the situation at the final generations (10,000) where $b$ is 0.3 and the tag number is 4. Interestingly, when the number of tags is increased, the intermediate $b$ values ($0.3 \leq b \leq 0.5$) yield more segregation.

**Subsection: Increasing neighborhood size**

The second extension was to increase the neighborhood size. The original model used a Moore neighborhood (i.e., 3 × 3 sites around a focal agent), which is represented as ns-1 in Fig. 7. We increased the neighborhood size to 5 × 5 sites, which is represented as ns-2 in the figure. By increasing the size and for lower values of $b$, $p_{Cs}$ decreases while $p_{Cd}$ increases, showing that it is harder for segregation to occur. If the neighborhood size is increased, agents can sense more further agents' tags. In this case, the sensitivity to the identical (or different) tags is decreased giving less relevance to the "meaning" of these tags. We thus found that small neighborhood size is better to maintain high level of communitarian cooperation and segregation.

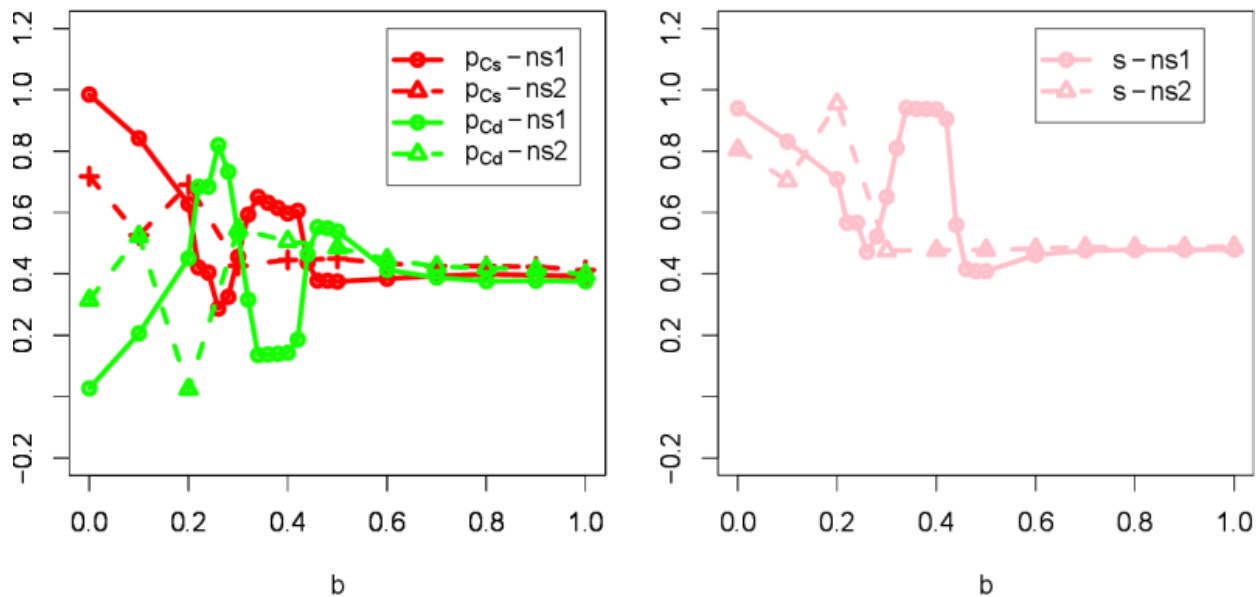

**Fig. 7.** Increasing neighborhood size. ns-1 means the original 3 × 3 sites centered on a focal agent (Moore neighborhood). ns-2 means the 5 × 5 sites as the extension. More spatially well-mixed configurations take place for any *b* value.

**Section: Discussion**

In this paper we have shown the necessary conditions for the emergence of high level of tag-based cooperation and segregation in an evolutionary spatial social game.

It is a first study allowing to assess the effect of all parameters needed for the emergence of cooperation and segregation, although some of the conditions have been previously covered. For example, Spector and Klein (2006) have shown, as we have, that lower mutation rates for the strategy promote tag mediated cooperation. For the coevolution of segregation and cooperation, such genetic stability is critical since tag-mediated cooperation frequently collapses in the case that a tag is not linked to a strategy.

In other studies, although it is already known that the combination of tag and spatial structure promotes cooperation, we showed that this effect is greatly enhanced when migration is also incorporated. This is because the identical tag groups are easily created by the movement of agents. Hammond et al. (2006) incorporated "immigrants", new agents showing up over time, but such immigrants could not move through the space. Since migration is a fundamental trait of animals and humans, we believe the situation described in this paper to be much more reminiscent of what happens in the real world.

The main original result of merging the migration and grouping strategy of Schelling together with the cooperative/defective evolution of May/Nowak is to facilitate a new route for cooperation favorable to agents which choose to move and to restrict their cooperative attitude to others sharing their same identity.

This communitary-restricted regrouping and cooperation is obviously well known to be a definitive salient trait of human nature. While the Schelling's model never really justified why even very tolerant agents could move to other random places, the simulation discussed in this paper shows that the cooperative gain and the increase in cooperative opportunities (against the prisoner dilemma defective trap) might be the real pressure force that encourages people to assemble according to some distinctive traits (color, religion, social classes, etc.).

One possible application of our model is to be used for collective actions of small autonomous robots. Think about land developments of remote areas by such robots. The robots have two modes: one is random movements and the other is cluster actions. Under normal conditions, each robot develops the land alone. However, the developments may sometimes need cooperative works of the robots. In that case, if the robots can use tag-like information, then they can gather efficiently to do the works.

In the present study, all traits of agents are free to evolve. Nevertheless, the co-evolution of communitarian segregation and cooperation was observed even for a wide range of parameters.